\documentclass{ws-ijmpe}

\begin{document}

\markboth{Authors' Names}{Instructions for  
Typing Manuscripts (Paper's Title)}

\catchline{}{}{}{}{}

\title{MICROSCOPIC DESCRIPTION OF NUCLEAR\\
WOBBLING MOTION\\
--- Rotation of triaxially deformed nuclei ---}

\author{TAKUYA SHOJI and \underline{YOSHIFUMI R. SHIMIZU}}

\address{Department of Physics, Graduate School of Sciences,\\
Kyushu University, Fukuoka 812-8581, Japan}

\maketitle

\begin{history}
\received{(received date)}
\revised{(revised date)}
\end{history}

\begin{abstract}
The nuclear wobbling motion in the Lu region is studied
by the microscopic cranked mean-field plus RPA method.
The Woods-Saxon potential is used as a mean-field with
a new parameterization which gives reliable description
of rapidly rotating nuclei.
The prescription of symmetry-preserving residual interaction
makes the calculation of the RPA step parameter-free,
and we find the wobbling-like RPA solution if the triaxial deformation
of the mean-field is suitably chosen.  It is shown that
the calculated out-of-band $B(E2)$ of the wobbling-like solution
depends on the triaxial deformation in the same way as
in the macroscopic rotor model, and can be used
to probe the triaxiality of the nuclear mean-field.

\end{abstract}

\section{Introduction}
\label{sec:intro}

In this talk we would like to discuss a fundamental question:
How nucleus rotates if it is triaxially deformed?
Namely, the question is whether nucleus exhibits the so-called
wobbling rotational motion.  We have been conducting research
on this subject for many years, and we would like to report,
especially, on the recent development of our microscopic study
of the wobbling rotation.

\begin{figure}[th]
\centerline{
\psfig{file=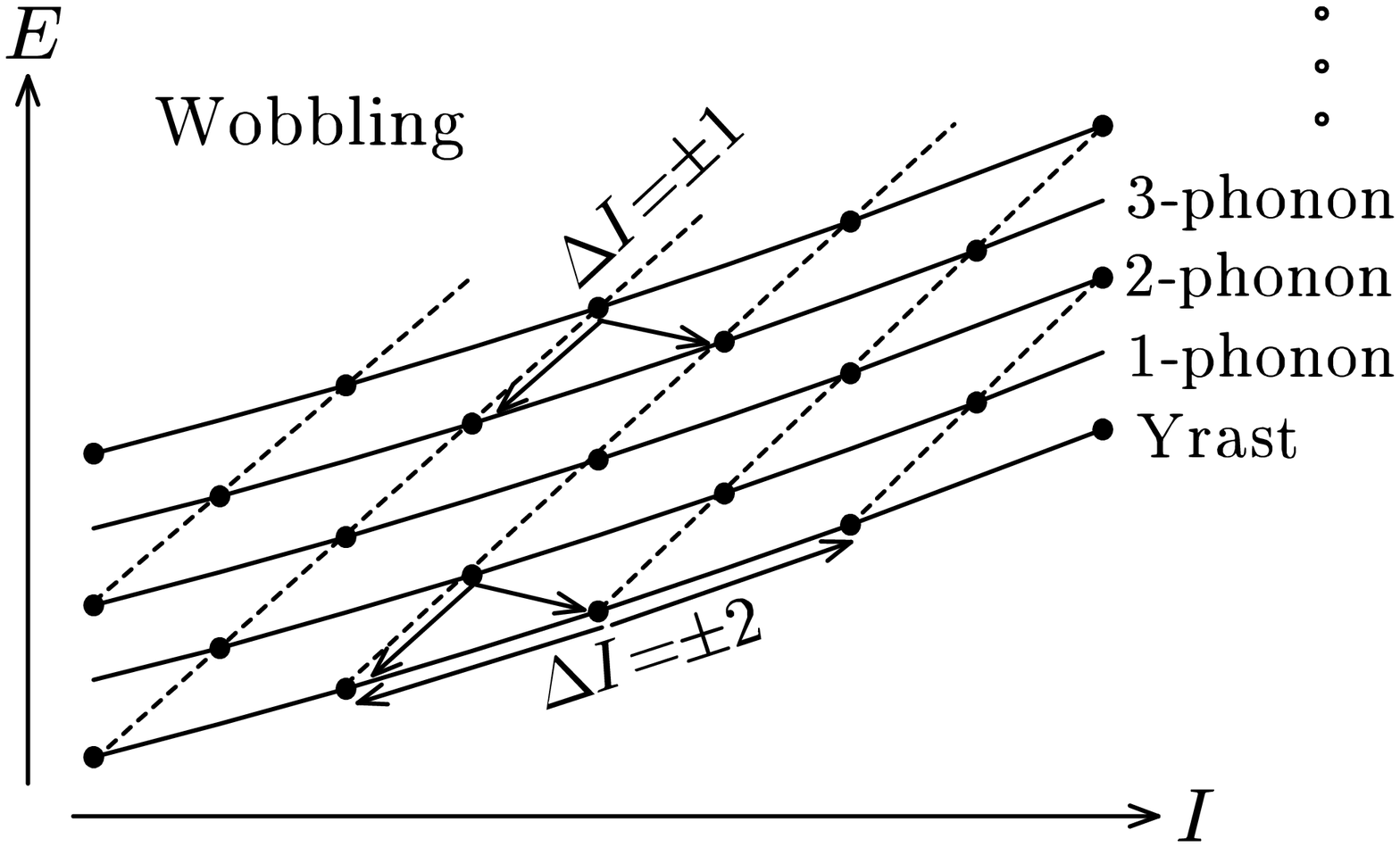,height=40mm}
\hspace*{2mm}
\psfig{file=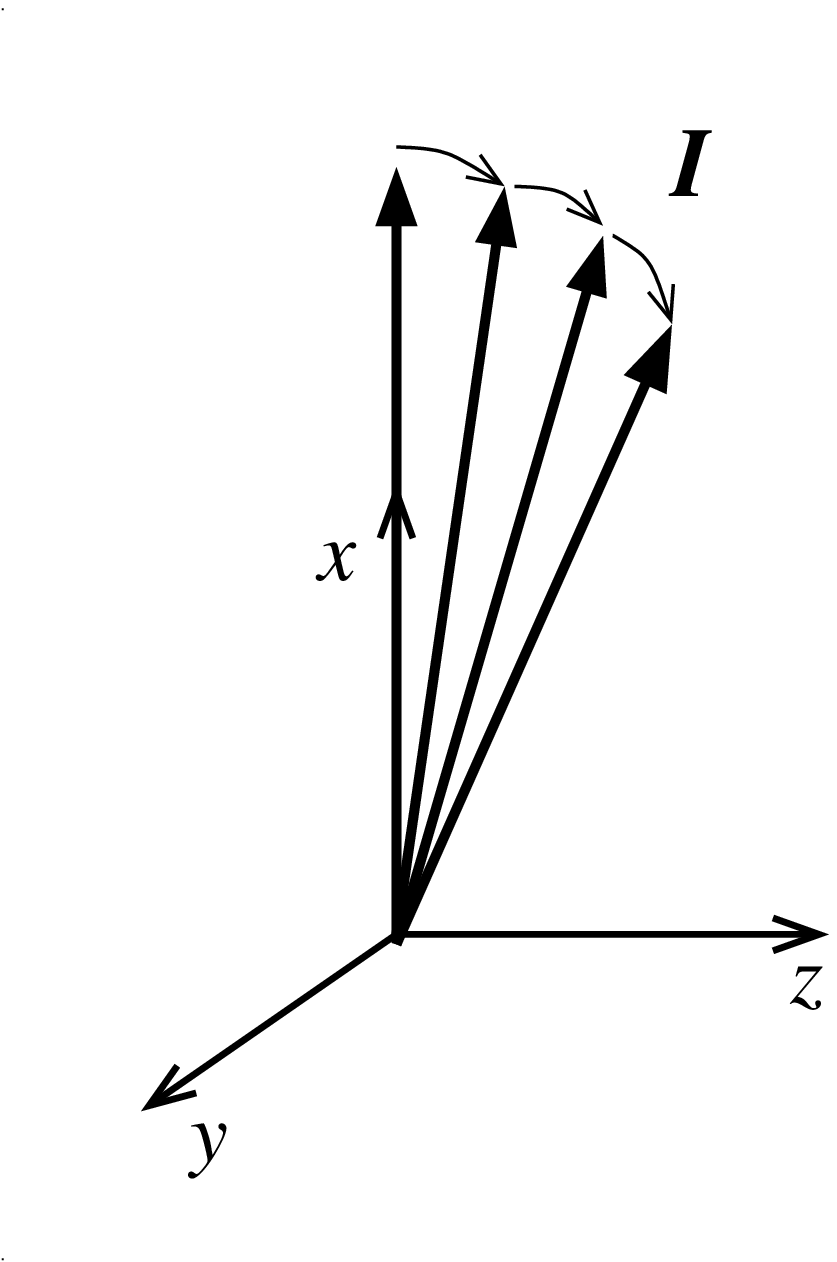,height=40mm}
}
\vspace*{-3mm}
\caption{A schematic illustration of the band structure of
the wobbling motion. }
\vspace*{-3mm}
\label{fig:wobs}
\end{figure}

The wobbling motion is an analogy of the quantized motion of the
macroscopic (``rigid-body'') triaxial rotor.
The spectra of the model hamiltonian of triaxial rotor can be solved
approximately (using the $\hbar=1$ unit) like this;\cite{BM75}
\begin{equation}
H_{\rm rot}=\frac{I_x^2}{{\cal J}_x}
+\frac{I_y^2}{{\cal J}_y}+\frac{I_z^2}{{\cal J}_z},
\hspace*{2mm}\Rightarrow\hspace*{2mm}
E(I,n) \approx\frac{I(I+1)}{2{\cal J}_x} +
\omega_{\rm wob}(n+1/2),
\label{eq:trot}
\end{equation}
where the wobbling energy $\omega_{\rm wob}$ is given by
\begin{equation}
\omega_{\rm wob}=\left(\frac{I}{{\cal J}_x}\right)\,
\sqrt{\left(\frac{{\cal J}_x}{{\cal J}_y}-1\right)
\left(\frac{{\cal J}_x}{{\cal J}_z}-1\right)}.
\label{eq:omwob}
\end{equation}
Here it is assumed that the main rotation axis is the $x$-axis
$({\cal J}_x>{\cal J}_y,{\cal J}_z)$.
This type of rotational motion is only possible when the system
is triaxially deformed, and appears as a multiple band structure
shown in Fig.~\ref{fig:wobs}.  It is composed of a group of
bands lying on top of each other, whose
horizontal sequences are usual rotational bands connected by the strong
stretched ($|{\mit\Delta}I|=2$) $E2$ transitions,
while the vertical sequences are connected
by slightly weaker $|{\mit\Delta}I|=1$ $E2$ transitions.
In this way, the vertical excitation, which corresponds to tilting
the angular momentum vector (in the body-fixed frame),
can be regarded as the phonon-like excitation,
$n=1,2,...$ in Eq.~(\ref{eq:trot}),
whose phonon energy is given by the famous formula~(\ref{eq:omwob}).

Until quite recently, there is no definite evidence of the multiple band
structure that exhibits this characteristic feature.  It has, however,
been first discovered in $^{163}$Lu:\cite{Od01} The one phonon wobbling
band has been measured nowadays in some Lu isotopes,\cite{Sch03,Amr03,Bri05}
and moreover
the two phonon wobbling band has been identified in $^{163}$Lu.\cite{Jen02}
These bands are examples of the so-called triaxial superdeformed (TSD) bands
in the Lu and Hf region.
The measured in-band (${\mit\Delta}I=-2$) $B(E2)$ values
between the horizontal sequences
in these TSD bands are typically about 500--700 Weisskopf units,
and the out-of-band (${\mit\Delta}I=-1$) $B(E2)$ values
from the one-phonon wobbling band
to the yrast TSD band are about 100 units or more.\cite{Gor04}
These values are very large and consistent to the wobbling picture
predicted by the macroscopic particle-rotor model,\cite{Ham02,Ham03}
where an odd $i_{13/2}$ quasiproton is coupled
to the rotor.

The triaxial deformation is crucial to realize the wobbling motion.
Then, it is important to know how the effects of triaxiality appear
in observables.  One is the $B(E2)$, which are related to
the two intrinsic quadrupole moments, $Q_{20}$ and $Q_{22}$,
\begin{equation}
Q_{20}=\sqrt{\frac{5}{16\pi}} \sum_{a=1}^A (2z^2-x^2-y^2)_a,
\quad
Q_{22}=\sqrt{\frac{15}{32\pi}} \sum_{a=1}^A (x^2-y^2)_a,
\label{eq:Qs}
\end{equation}
and the triaxiality parameter $\gamma$ is defined in terms of them
as usual:
\begin{equation}
\tan\gamma=-\frac{\sqrt{2}\,\langle Q_{22}\rangle}{\langle Q_{20} \rangle}
\quad \hbox{(Lund convention of sign)}.
\label{eq:gamdef}
\end{equation}
The other is the energy spectra, which reflects the properties of
the three different moments of inertia about intrinsic axes.
Here, it should be pointed out that
the characteristic feature of the out-of-band $E2$ transitions
observed in Lu nuclei suggests the ``positive $\gamma$'' shape,
i.e. the main rotation axis (the $x$-axis in our notation) is the shortest
axis in that shape.  This shape completely contradicts the well-known
irrotational moments of inertia, in which ${\cal J}_y$ is the largest,
and then the wobbling frequency~(\ref{eq:omwob}) becomes imaginary.
Therefore, we don't know what kind of inertia should be
used in macroscopic models, and a microscopic approach is necessary.

\section{Microscopic approach}
\label{sec:mic}

In order to understand the wobbling motion in Lu nuclei, we use
a standard microscopic approach;
the cranked mean-field and the random phase approximation (RPA).
The basic idea is to describe the horizontal rotational bands
in Fig.~\ref{fig:wobs} by the semiclassical cranking prescription,
and at the same time, the vertical phonon-like excitation in terms of
the RPA, which is known to be successful for describing the collective
vibrational modes.  Namely,
\begin{eqnarray}
 && h'=h_{\rm def} -\omega_{\rm rot} J_x,
 \quad \hbox{(cranking)}
\label{eq:crank}\\
 && [H',X_n^\dagger]=\omega_n X_n^\dagger, 
 \quad \hbox{(RPA) \quad with } H'=h'+V_{\rm int},
\label{eq:RPA}
\end{eqnarray}
where $h_{\rm def}$ is the triaxially deformed mean-field hamiltonian
with the cranking frequency $\omega_{\rm rot}$ about the $x$-axis,
$X_n^\dagger$ is the creation operator of the $n$-th RPA eigen mode
with the eigen energy $\omega_n$, and $V_{\rm int}$ is the residual
interaction.

It should be emphasized that the residual interaction is constructed
so as to restore the rotational symmetry broken
by a general mean-field, i.e.
\begin{equation}
 [h_{\rm def},J_k]\ne 0
 \quad\Rightarrow\quad
 [h_{\rm def}+V_{\rm int},J_k]=0,\quad (k=x,y,z),
\label{eq:symres}
\end{equation}
with
\begin{equation}
  V_{\rm int}=-\frac{1}{2}\sum_{k=x,y,z}\kappa_k F_k^2,\quad
 F_k\equiv[h_{\rm def},iJ_k],\,\,\,
 \kappa_k \equiv \langle [[h_{\rm def},J_k],J_k]\rangle,
\label{eq:Vres}
\end{equation}
where the expectation value is taken with respect to the cranked
mean-field yrast state.
Thus the RPA solutions are determined without any
ambiguities by a given mean-field, and there are no adjustable
parameters in the RPA step.

\begin{figure}[hbt]
\centerline{\psfig{file=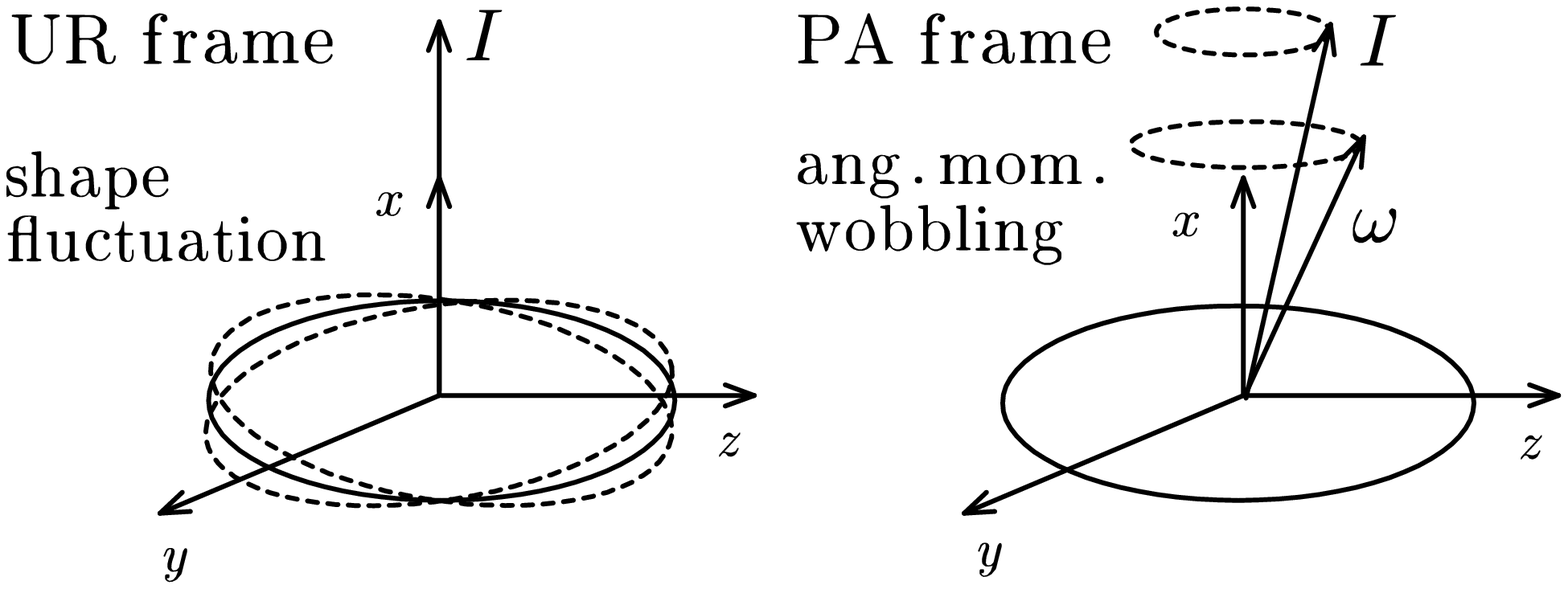,height=35mm}}
\vspace*{-3mm}
\caption{
A schematic illustration depicting the relation between
two dynamical pictures in the uniformly rotating (UR) frame and
the principal axis (PA) frame.  }
\vspace*{-3mm}
\label{fig:wobpic}
\end{figure}

However, it is not clear how the result of such a microscopic
approach is related to the wobbling picture of the macroscopic rotor model.
It was shown by Marshalek\cite{Mar79} many years ago that,
if an appropriate RPA mode exists, the rotor model wobbling picture
naturally appears by going from the uniformly rotating (UR) frame,
where the cranking prescription is applied,
over to the principal axis (PA) frame.
Taking into account the fact that the RPA can be regarded as a small
amplitude limit of the time-dependent mean-field theory,
the time-dependent mean-field describing the wobbling motion
is given in the UR frame as
\begin{equation}
 h_{\rm UR}(t)=h_{\rm def} - \omega_{\rm rot} J_x
  - \kappa_y {\cal F}_y(t)\,F_y- \kappa_z {\cal F}_z(t)\, F_z,
\label{eq:hUR}
\end{equation}
where ${\cal F}_k(t)\equiv \langle F_k \rangle_{\rm UR}(t)$ ($k=y,z$)
are the UR frame expectation values; only the $y$ and $z$ components
are relevant to the wobbling phonon excitation,
which transfers spin by ${\mit\Delta}I=\pm 1$.
In the case of the harmonic oscillator potential,
e.g. the Nilsson potential, the operators $F_y$ and $F_z$
are proportional to $Q_y$ and $Q_z$,
the non-diagonal quadrupole tensor operators.
Even in the case of general potentials,
these non-diagonal parts $Q_k$ ($k=x,y,z$) are used to define the PA frame;
\begin{equation}
\langle Q_k \rangle_{\rm PA} =0, \,\,\hbox{(PA condition)},
\quad Q_k=\sqrt{\frac{15}{4\pi}}\sum_{a=1}^A (x_i x_j)_a,\,\,
(k,i,j)\hbox{-cyclic}.
\label{eq:Qnon}
\end{equation}
The shape fluctuation vanishes by this PA frame condition,
but the fluctuation of the angular momentum vector appears in place of it;
thus the time-dependent mean-field in the PA frame is transformed as
\begin{equation}
 h_{\rm PA}(t)=h_{\rm def} - \omega_x(t) J_x
  - \omega_y(t) J_y - \omega_z(t) J_z,
\label{eq:hPA}
\end{equation}
where $\omega_x(t) \approx \omega_{\rm rot}$ in the small amplitude limit,
and the presence of $\omega_y(t)$ and $\omega_z(t)$ reflects
that the angular frequency vector also fluctuates
about the main rotation axis (the $x$-axis).
The relation between the UR and PA frames is depicted schematically
in Fig.~\ref{fig:wobpic}.
The PA frame corresponds to the body-fixed frame of the rotor,
and the moments of inertia are naturally introduced by
\begin{equation}
 {\cal J}_x\equiv\langle J_x \rangle/\omega_{\rm rot}, \quad
 {\cal J}_y(n)\equiv J_y(n)/\omega_y(n),\,\,\,
 {\cal J}_z(n)\equiv J_z(n)/\omega_z(n),
\label{eq:Tmom}
\end{equation}
where $J_k(n)$ and $\omega_k(n)$ ($k=y,z$) are Fourier components
of the PA frame expectation values
with respect to the $n$-th RPA eigen mode.
Using these microscopically calculated moments of inertia,
the RPA excitation energy can be written in exactly
the same form~(\ref{eq:omwob}) as in the macroscopic rotor model.
It can be also shown\cite{SM95} that the out-of-band
$B(E2)$ calculated by the RPA approach can be expressed
in the same way as in the rotor model.

\section{Results of microscopic calculations}
\label{sec:res}

Realistic calculations based on the Marshalek's theory have been
carried out in the previous works.\cite{MM90,SM95}
After the discovery of the wobbling phonon excitation in the TSD bands
in the Lu and Hf region, new calculations have been performed,
and it is confirmed that the wobbling-like RPA solutions
do exist.\cite{MSM02,MSM04}
However, the Nilsson potential has been used in these calculations,
and they are suffered from the problem of large moments of inertia
because of the spurious velocity dependence coming from
the {\boldmath $l$}$^2$-term.
Therefore, we have conducted new research with using
the Woods-Saxon potential as a nuclear mean-field.
Especially, a new parameterization of the Woods-Saxon potential
has been provided quite recently by Ramon Wyss,
with which it is possible to nicely reproduce
the basic properties like the neutron and proton density distributions.
This is very important to obtain reliable results for the moments of inertia
and the quadrupole moments.

\subsection{Mean-field parameters}
\label{subsec:mean}

\begin{figure}[hbt]
\centerline{
\psfig{file=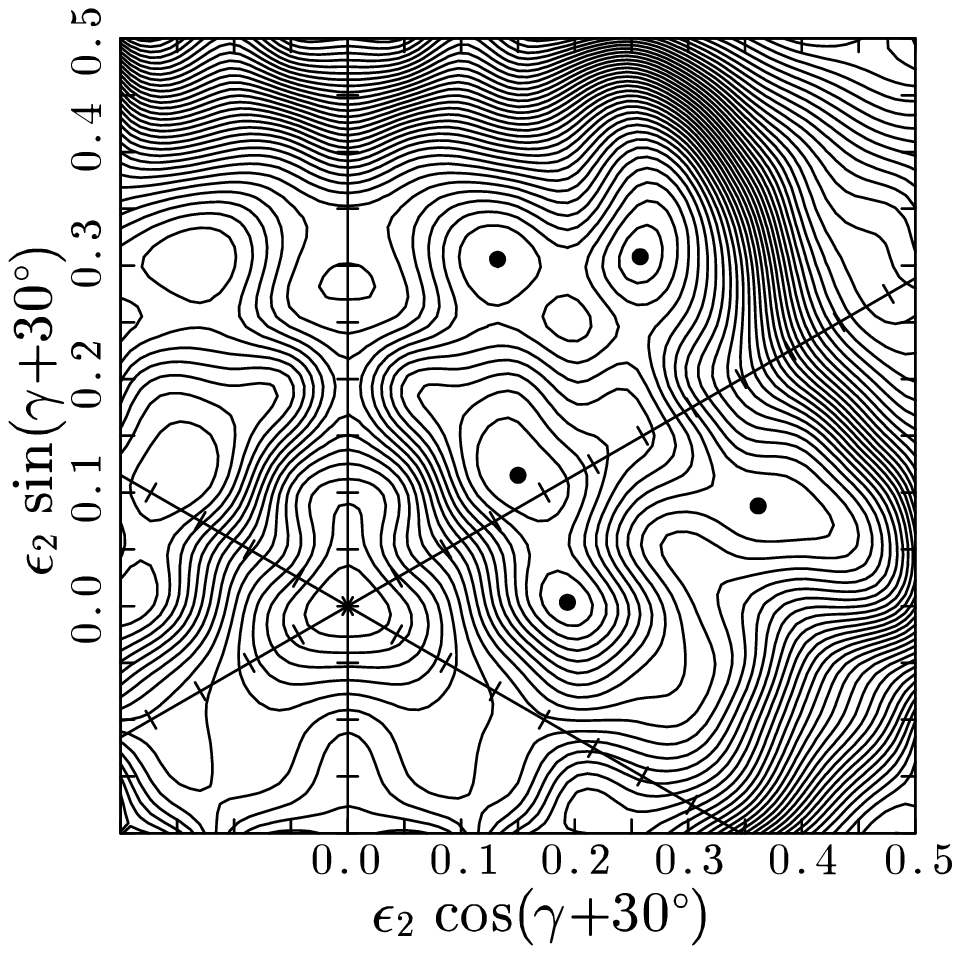,height=55mm}
\hspace*{2mm}
\psfig{file=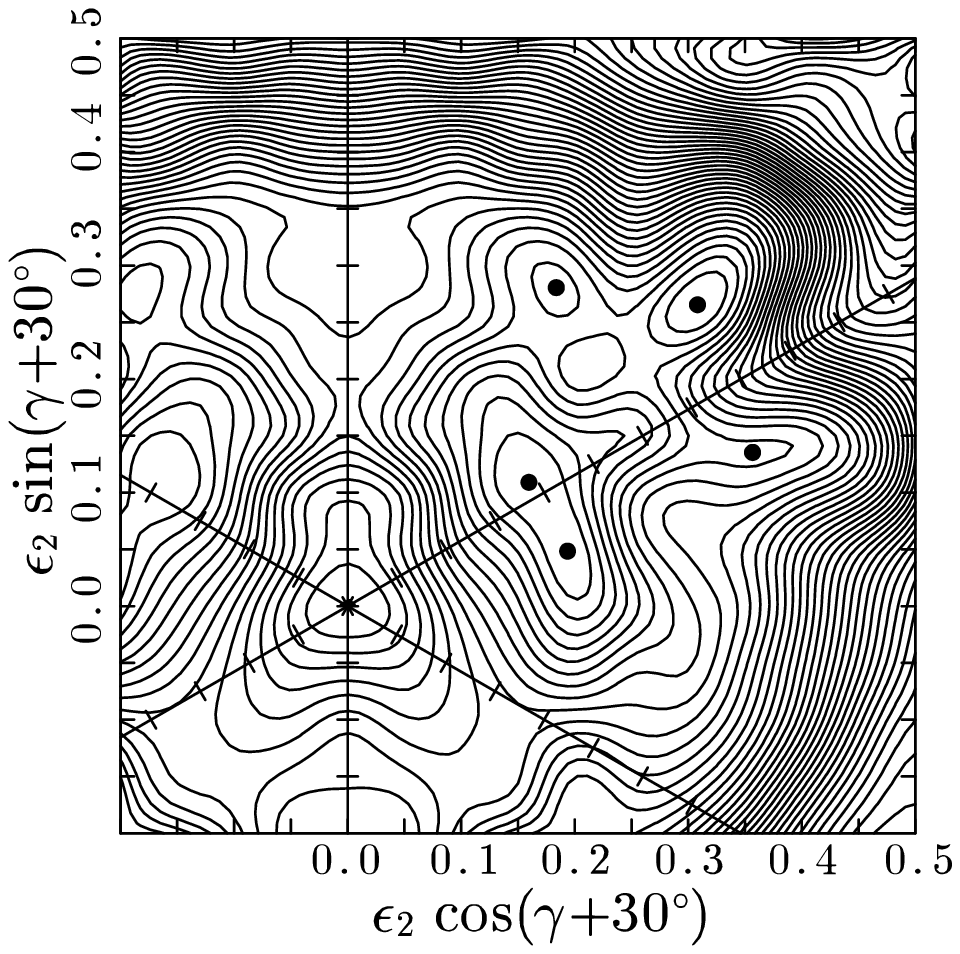,height=55mm}
}
\vspace*{-3mm}
\caption{
An example of the potential energy surface obtained by
the cranked Nilsson Strutinsky calculation for
the $\pi i_{13/2}$ $(\pi,\alpha)=(+,+1/2)$ configuration
at $I=53/2^+$ in $^{163}$Lu.
The energy between contours is 250 keV.
In the left figure is adopted the ``usual'' $\gamma$,
which is used to parameterize the Nilsson potential,
while in the right one the calculations are same but
the $\gamma$ defined by a similar Eq. to~(\ref{eq:gamdef})
(see the footnote in the text) is adopted.}
\vspace*{-3mm}
\label{fig:Nstr}
\end{figure}

As it is discussed in the previous section, there is no adjustable
parameter in the RPA step once the mean-field is specified.
Therefore, let us briefly explain the mean-field parameters used
in our calculation.  In principle, they should be determined
selfconsistently by minimizing the energy.
Actually we are developing the Woods-Saxon Strutinsky calculations,
but it is not available yet.  Therefore we have taken the deformation
parameters corresponding to the minimum
of the Nilsson Strutinsky calculations, an example of which
is shown in Fig.~\ref{fig:Nstr}.
The TSD minimum moves slightly as a function of the rotational frequency
but the amount of change is small, so that we have used average values,
$\beta_2=0.42$, $\gamma=12^\circ$, $\beta_4=0.034$ in all the calculations.
we have checked that the qualitative feature of the calculated results
does not change if these values are modified in a reasonable range.

Here we would like to point out that there are various different
definitions for the triaxiality parameter $\gamma$.
In this presentation,
we use the one defined by Eq.~(\ref{eq:gamdef}) throughout,
which is directly related to the two intrinsic quadrupole moments,
and so the $B(E2)$.
In Fig.~\ref{fig:Nstr}, the same calculations are employed
but two representations with different definitions of $\gamma$
are shown.  The left part is an usual representation,
where the $\gamma$ is the one used in the Nilsson potential,
and the TSD minimum has $\gamma\approx 20^\circ$ and $\epsilon_2\approx 0.4$.
In the right part the definition similar to Eq.~(\ref{eq:gamdef})
is used\footnote{
 More precisely, the expectation values in Eq.~(\ref{eq:gamdef}) is
 replaced by those with respect to the liquid-drop like
 sharp cut-off density distribution.
 Then the $\gamma$ is determined purely by the geometry
 of the potential without recourse to actual wave functions. }
keeping the $\epsilon_2$ as usual,
and then the TSD minimum has $\gamma\approx 12^\circ$.
Thus, the difference between the two definitions for the same shape
is rather large for large $\epsilon_2$ deformation like in the case
of the TSD bands.
More details about the various definitions of $\gamma$ parameters
and relations between them will be discussed elsewhere.\cite{SS06}

\begin{figure}[hbt]
\centerline{ \psfig{file=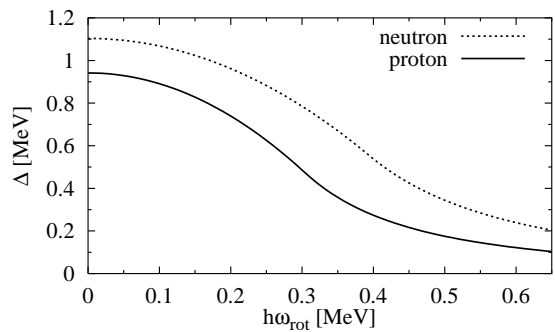,height=45mm} }
\vspace*{-3mm}
\caption{The neutron and proton gap parameters used in
the calculation for the TSD band in $^{163}$Lu.}
\vspace*{-3mm}
\label{fig:GAP}
\end{figure}

The neutron and proton pairing gaps are also important mean-field
parameters.  We have used the following parameterization, which is
convenient to avoid the abrupt pairing collapse;
\begin{equation}
{\mit\Delta}(\omega_{\rm rot})={\mit\Delta}_0\times\left\{\begin{array}{lc}
\left[1-
{\displaystyle \frac{1}{2}\left(
\frac{\omega_{\rm rot}}{\omega_{\rm c}}\right)^2}\right],
& \quad\omega_{\rm rot} < \omega_{\rm c}, \\
\quad {\displaystyle \frac{1}{2}\left(
\frac{\omega_{\rm c}}{\omega_{\rm rot}}\right)^2},
& \quad\omega_{\rm rot} \ge \omega_{\rm c},
\end{array}\right.
\label{eq:Delpar}
\end{equation}
where ${\mit\Delta}_0$ is given by the even-odd mass differences
of neighboring even-even nuclei, and $\omega_{\rm c}$ is
determined by the selfconsistent monopole pairing calculation.
The resultant pairing gap parameters are shown in Fig.~\ref{fig:GAP}.
Now all the mean-field parameters are fixed, and
there is no adjustable parameter for the following RPA calculations.

\subsection{RPA calculations}
\label{subsec:rpa}

\begin{figure}[hbt]
\centerline{
\psfig{file=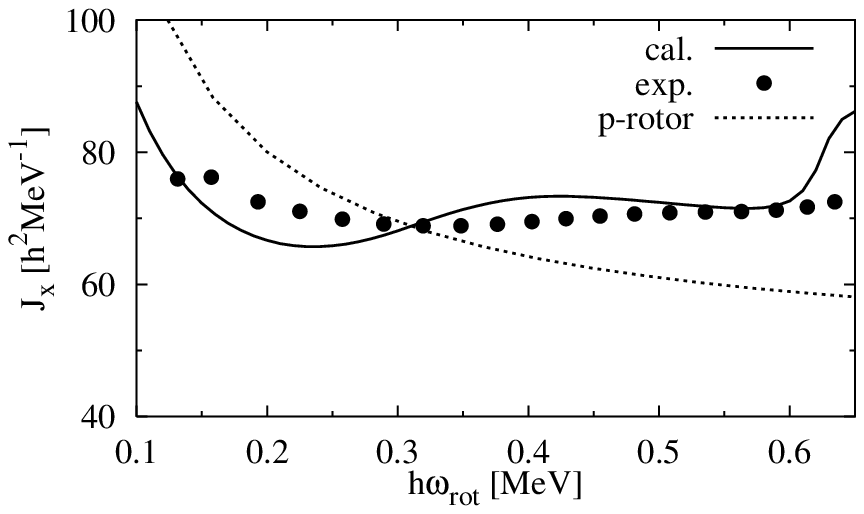,height=40mm}
\hspace*{3mm}
\psfig{file=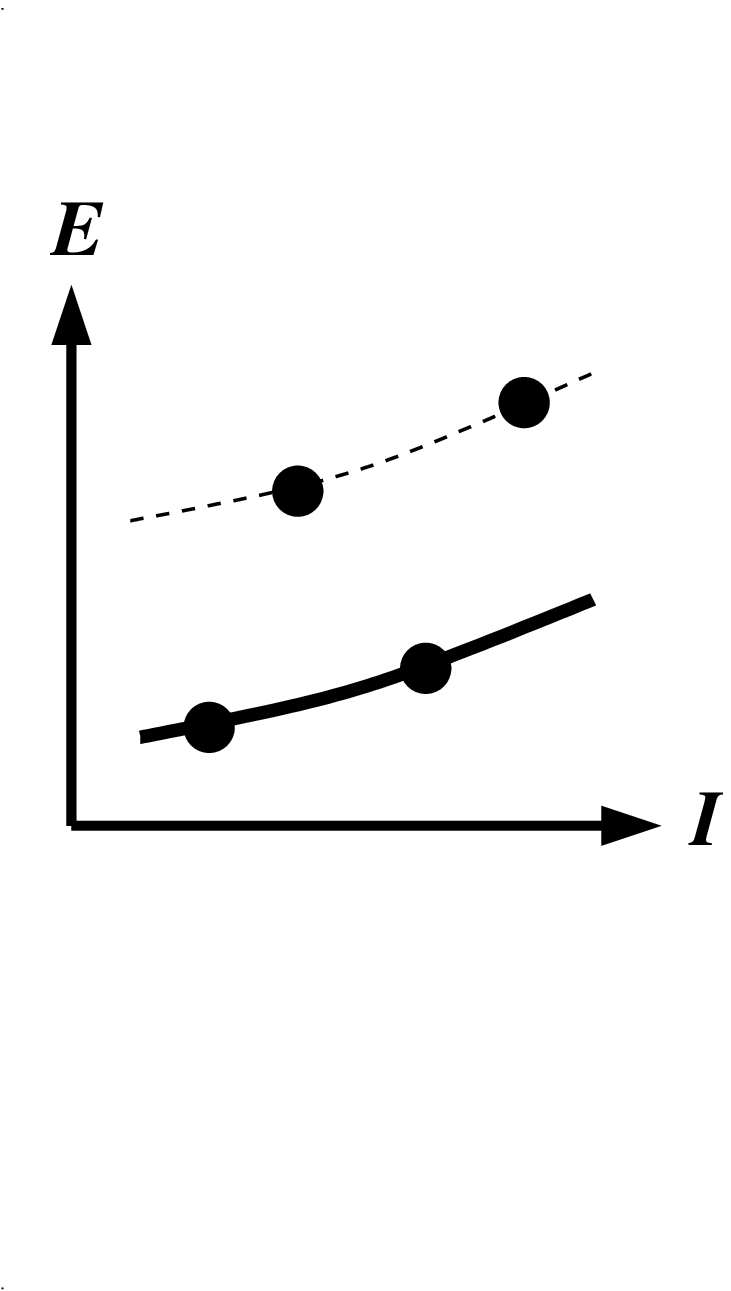,height=40mm}
}
\vspace*{-3mm}
\caption{The moment of inertia, ${\cal J}_x$ in Eq.~(\ref{eq:Tmom})
as a function of the rotational frequency in $^{163}$Lu.
The result of the particle-rotor model is also included
as the dotted line with the legend ``p-rotor''.}
\vspace*{-3mm}
\label{fig:JX}
\end{figure}

Now let us present the results of RPA calculations for the TSD
wobbling band in $^{163}$Lu,
for which most extensive data are available.\cite{Gor04,Jen02n,Jen04}
First, Fig.~\ref{fig:JX} shows the cranking moment of inertia,
${\cal J}_x$ in Eq.~(\ref{eq:Tmom}), as a function of the rotation frequency.
This inertia is related to the slope of
the horizontal rotational band in Fig.~\ref{fig:wobs}.
Our calculation agrees the trend of experimental data rather well.
As for a reference, here and in the following, we also include
examples of the particle-rotor model calculations
by Hamamoto-Hagemann;\cite{Ham03} the relevant parameters of the model are
$\gamma=20^\circ$, and
${\cal J}_x^{\rm (R)},{\cal J}_y^{\rm (R)},{\cal J}_z^{\rm (R)}=48, 45, 17$
$\hbar^2$MeV$^{-1}$, respectively.
The result of the particle-rotor model is monotonically decreasing,
which can be understood by the following approximation
valid for a highly aligned quasiparticle band;
${\cal J}_x \approx {\cal J}_x^{\rm (R)}+j/\omega_{\rm rot}$, where
$j \approx 13/2$ is the maximal alignment of the $\pi i_{13/2}$ orbit.

\begin{figure}[hbt]
\centerline{
\psfig{file=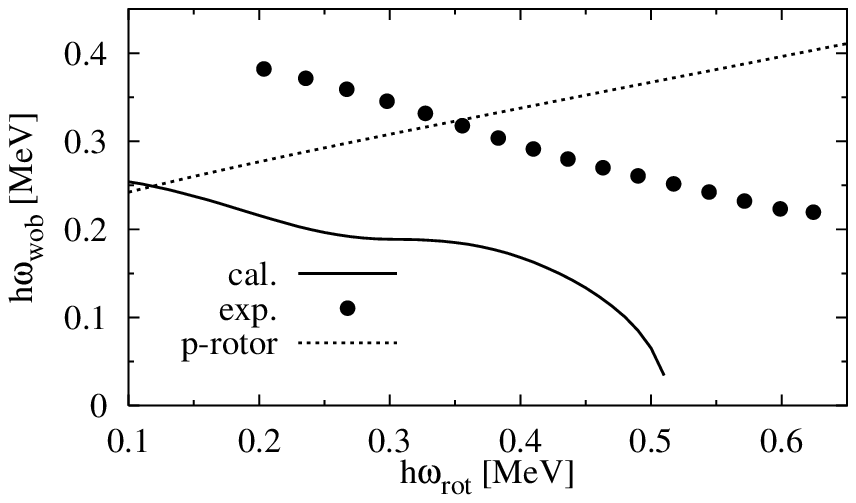,height=45mm}
\hspace*{3mm}
\psfig{file=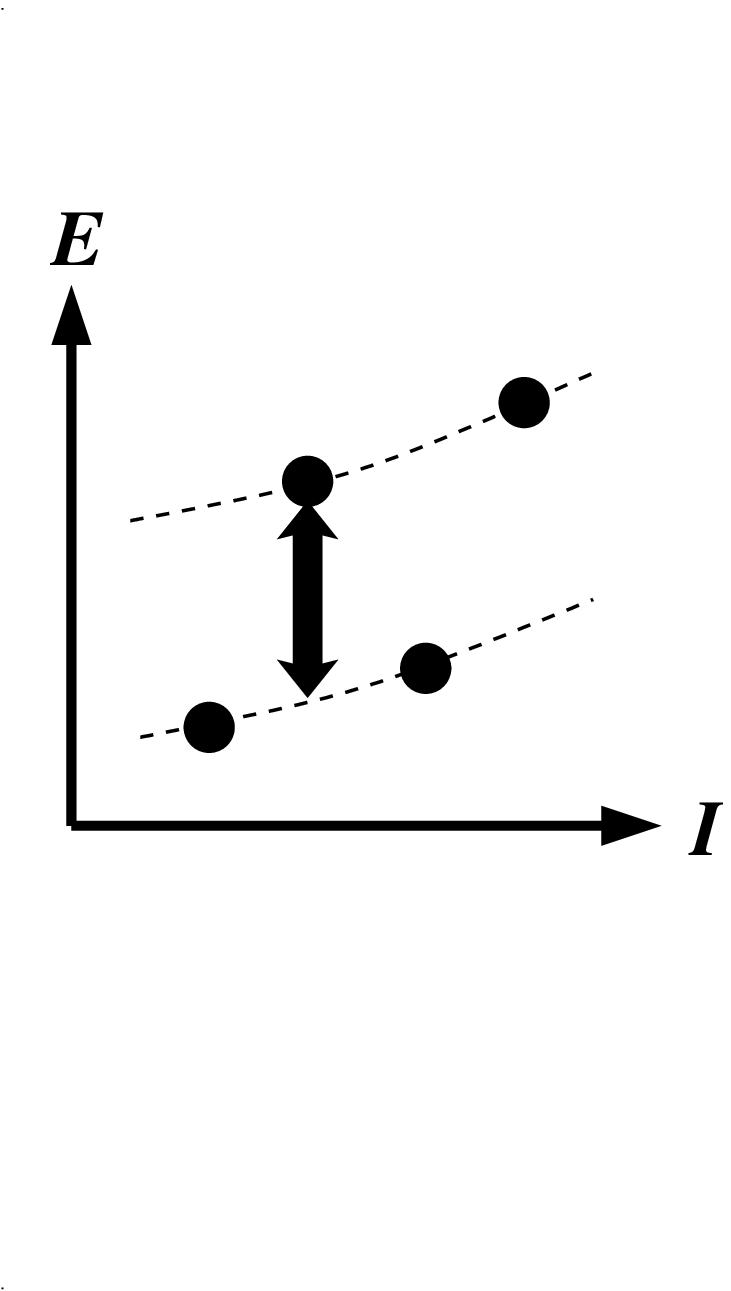,height=45mm}
}
\vspace*{-3mm}
\caption{The excitation energy of the one-phonon wobbling band
as a function of the rotational frequency in $^{163}$Lu.}
\vspace*{-3mm}
\label{fig:EN}
\end{figure}

Next we show the results of the RPA.
We have found that the wobbling-like RPA solution does exist
also for the Woods-Saxon mean-field potential;
the calculated excitation energy
is compared with experimental data in Fig.~\ref{fig:EN}.
The solution is stable against the change of mean-field parameters
in a reasonable range, so we believe that the existence of the wobbling-like
solution has been confirmed.
Although the result qualitatively agrees,
the calculated excitation energy is smaller than the measured one,
and it vanishes at about $\omega_{\rm rot}=0.52$ MeV.
For the excitation energy, the particle-rotor model gives
increasing energy as a function $\omega_{\rm rot}$ in contrast to the data.
It is, however, difficult to change this general trend as long as
the constant moments of inertia of the rotor are used,
see Eq.~(\ref{eq:omwob}).

\begin{figure}[hbt]
\centerline{ \psfig{file=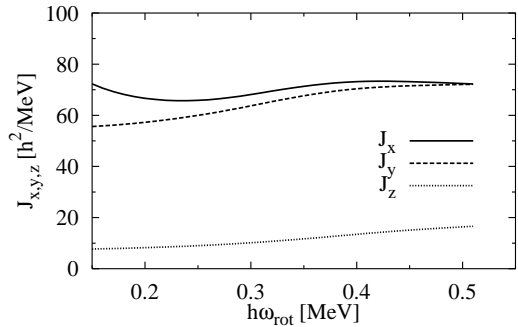,height=45mm} }
\vspace*{-3mm}
\caption{There moments of inertia calculated by the Marshalek's theory,
Eq.~(\ref{eq:Tmom}), applied to the wobbling-like RPA solution in $^{163}$Lu.}
\vspace*{-3mm}
\label{fig:J3}
\end{figure}

In Fig.~\ref{fig:J3}, we show three moments of inertia
calculated  according to the Marshalek's theory by Eq.~(\ref{eq:Tmom}).
Note that they are not the inertia of the rotor, but the total inertia
of the system; it is not possible to divide the contribution
into the particle and the rotor parts {\it a priori}
in the microscopic calculation.
They are not constant but gradually change, which is necessary to
understand the decreasing trend of the observed excitation energy.

\begin{figure}[hbt]
\centerline{
\psfig{file=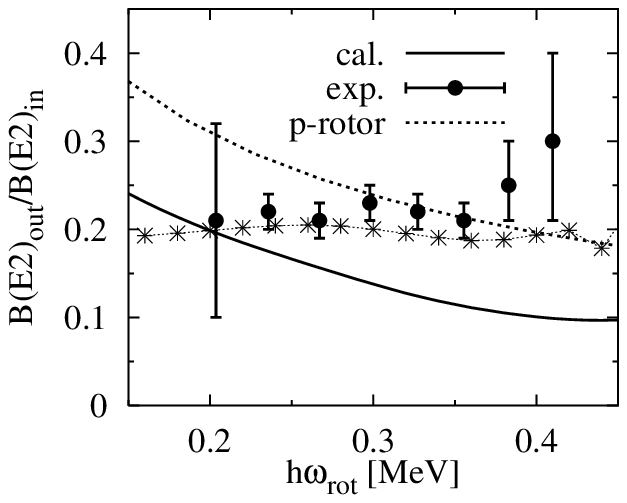,height=45mm}
\hspace*{1mm}
\psfig{file=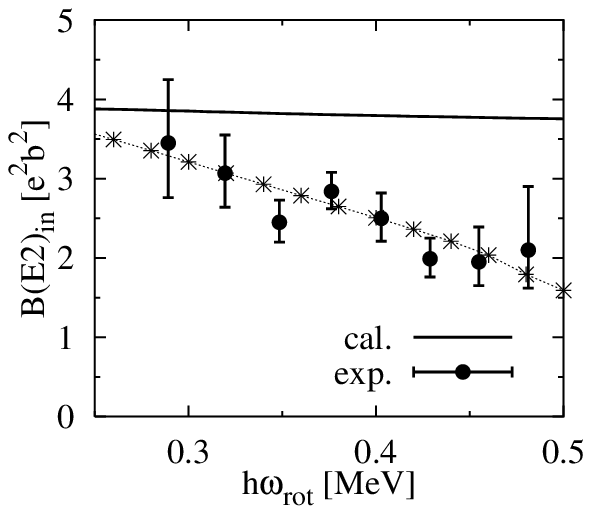,height=45mm}
}
\vspace*{-3mm}
\caption{The out-of-band over in-band $B(E2)$ ratio (left) and
the in-band $B(E2)$ (right)
as a function of the rotational frequency
for the one-phonon wobbling band in $^{163}$Lu.
The lines connected with star symbols 
are results of the specific RPA calculation,
where the mean-field parameter $\gamma$ is changed
from $\gamma=12^\circ$ at $\omega_{\rm rot}=0.2$ MeV
to $\gamma=22^\circ$ at $\omega_{\rm rot}=0.4$ MeV.
}
\vspace*{-3mm}
\label{fig:BE2}
\end{figure}

Now we come to one of the most important observables, the $B(E2)$,
which are depicted in Fig.~\ref{fig:BE2}.
The measured out-of-band over in-band $B(E2)$ ratio is
almost constant (the left part of Fig.~\ref{fig:BE2}), but
a gradual decrease of the $B(E2)$ ratio 
as a function of $\omega_{\rm rot}$ is expected
in the rotor model if the triaxiality parameter $\gamma$ is kept constant.
Both ratios calculated by the RPA approach and by the particle-rotor model
follow this tendency, but our microscopic result
is about 60\% of the measured value at $\omega_{\rm rot}\approx 0.3$ MeV,
where the particle-rotor model gives correct magnitude.
However, the calculated out-of-band $B(E2)$ values are more than
50 Weisskopf units, which is a huge value for a normal RPA solution,
and clearly indicates that the obtained solution is not of vibrational
but of rotational character ($B(E2)$ values of typical collective
vibrations in deformed nuclei are about 10 Weisskopf units at most).
The average values of the $B(E2)$ ratio in the particle-rotor model agrees 
rather well; in fact the parameter $\gamma=20^\circ$ is chosen
for this reason.\cite{Ham03}  Note that our mean-field value is
$\gamma=12^\circ$, and is much smaller; we have checked that almost
the same larger value as in the particle-rotor model can be obtained
if we use a similar $\gamma$ value.  As for the in-band $B(E2)$,
it is just related to the quadrupole moment of the mean-field
about the rotating axis ($x$-axis).
The calculated result (the right part of Fig.~\ref{fig:BE2})
is almost constant because we have used constant deformation
parameters, while the data show a clear tendency to decrease.

As is discussed already, the measured $B(E2)$ strongly suggest
that the deformation of the mean-field, especially the triaxiality $\gamma$,
is changing as a function of the rotational frequency or spin.
In the simple rotor model, both the in-band $B(E2)$ and the $B(E2)$ ratio
can be estimated approximately as
\begin{eqnarray}
&&B(E2)_{\rm in} \approx \frac{15}{32\pi}e^2
 \langle \sum_\pi (y^2-z^2)\rangle^2
 =\frac{5}{32\pi}e^2{Q_\pi^2}
 \cos^2(\gamma+30^\circ),
\label{eq:RotBE2in}\\
&&\frac{B(E2)_{\rm out}}{B(E2)_{\rm in}} \approx
\frac{2}{I}\left[
\frac{(w_z/w_y)^{1/4}\sin(\gamma+60^\circ)
  +(w_y/w_z)^{1/4}\sin\gamma}{\cos(\gamma+30^\circ)}\right]^2,
\label{eq:RotBE2r}
\end{eqnarray}
with
\begin{equation}
 w_y\equiv 1/{\cal J}_z - 1/{\cal J}_x,\quad
 w_z\equiv 1/{\cal J}_y - 1/{\cal J}_x.
\end{equation}
The in-band $B(E2)$ depends on both the magnitude of deformation, e.g.
$\epsilon_2$, and the triaxiality $\gamma$, while the $B(E2)$ ratio
depends only on $\gamma$ in this approximation.  Moreover,
these expressions clearly shows that the $B(E2)$ are quite sensitive
to the triaxiality parameter $\gamma$.
The $1/I$ factor in Eq.~(\ref{eq:RotBE2r}) gives the decreasing trend
of the $B(E2)$ ratio if $\gamma$ and ${\cal J}_x,{\cal J}_y,{\cal J}_z$
are kept constant.
Now, can we understand the measured spin dependence of $B(E2)$
by changing the $\gamma$ parameter?
We have tried to play a game by increasing $\gamma$ linearly
from $\gamma=12^\circ$ at $\omega_{\rm rot}=0.2$ MeV
to $\gamma=22^\circ$ at $\omega_{\rm rot}=0.4$ MeV,
with keeping other parameters unchanged.
The results are shown by the lines with star symbols in Fig.~\ref{fig:BE2}.
Both the $B(E2)$ ratio and the in-band $B(E2)$ nicely agree with data.

\section{Summary}
\label{sec:sum}

We have performed microscopic RPA calculations using
the Woods-Saxon potential as a mean-field.  We have found that
the wobbling-like eigen mode does exist in this calculation.
Considering that the similar solutions have been obtained
in the previous investigation using
the Nilsson potential,\cite{MSM02,MSM04}
we believe that this confirms the existence of the collective mode,
the nuclear wobbling motion, suggested by the macroscopic rotor model.

The calculated excitation energy is smaller than the experimental
one by about 100--150 keV.  This may suggest that some
part of coupling effects to the odd $i_{13/2}$ quasiproton
is missing in our calculation.
In this respect, Hamamoto-Hagemann has discussed,\cite{Ham03}
by using a simple approximation in the particle-rotor model,
that the effect of coupling to the odd particle
appears as a constant energy shift to the wobbling energy~(\ref{eq:omwob}),
${\mit\Delta}\omega_{\rm wob}=j/{\cal J}_x$,
where $j$ is the quasiparticle alignment.
Applying this estimate to the case of $^{163}$Lu, $j=13/2$ and
${\cal J}_x \approx 70$ $\hbar^2$MeV$^{-1}$, the shift amounts to
${\mit\Delta}\omega_{\rm wob}\approx 93$ keV.
This fills most part of the energy underestimated in our calculation,
although we are not so sure whether such an energy shift
can be justified in our microscopic framework:
An explicit particle-vibration coupling treatment based on our
framework is necessary in order to elucidate this point.

The experimental out-of-band over in-band $B(E2)$ ratio on average
suggests larger triaxiality $\gamma\approx 20^\circ$ than the value
obtained for TSD minima in the Nilsson Strutinsky calculation,
$\gamma\approx 12^\circ$; there is an apparent discrepancy.
Note that $\gamma$ here is defined by Eq.~(\ref{eq:gamdef}),
and not the ``usual'' $\gamma$ in the Nilsson potential,
whose translated value is incidentally about $20^\circ$
in the case of the TSD bands.

Both the in-band $B(E2)$ and the $B(E2)$ ratio indicate an increase
of the triaxial deformation as a function of the rotational frequency or spin.
We can obtain nice agreements between the microscopic RPA calculations
and the experimental data, if the triaxiality parameter $\gamma$
is changed linearly, e.g. from $\gamma=12^\circ$ to $22^\circ$.
However, the Nilsson-Strutinsky calculation suggests
almost constant $\gamma$ values for TSD minima,
and it may be difficult to expect such a large change of $\gamma$ values.
We definitely need more study for quantitative understanding
of the observed properties of the nuclear wobbling motion
in the Lu region.

\end{document}